\documentclass[useAMS,usenatbib,usegraphicx,twocolumn]{mn2e}
  
\usepackage{aas_macros}
\usepackage{threeparttable}
\usepackage{grffile}
\usepackage{amssymb}
\usepackage{multirow}
\usepackage{wallpaper}
\usepackage{mathbbol}
\usepackage{rotating}
\usepackage{blindtext}

\usepackage{tocloft}
\newcommand{\fat}{\textbf}
\newcommand{\ita}{\textit}

\newcommand{\beq}{\begin{equation}}
\newcommand{\eeq}{\end{equation}}  
\newcommand{\RNum}[1]{\uppercase\expandafter{\romannumeral #1\relax}} 
 
\newcommand{\bp}{$b$ }        
\defcitealias{Koertgen15}{\scshape KB15}
\defcitealias{Koertgen16}{\scshape K16}
\title[Turbulent driving in MCs]{The driving of turbulence in simulations of molecular cloud formation and evolution}
  \author[B. K\"ortgen]
  {Bastian~K\"ortgen$^{1}$\thanks{bkoertgen@hs.uni-hamburg.de}, Christoph~Federrath$^{2}$, and Robi~Banerjee$^{1}$\\
  $^{1}$ Hamburger Sternwarte, Universit\"at Hamburg, Gojenbergsweg 112, 21029 Hamburg, Germany \\
  $^{2}$ Research School of Astronomy and Astrophysics, Australian National University, Canberra, ACT 2611, Australia\\
  }

\date{Released 2016}

\pagerange{\pageref{firstpage}--\pageref{lastpage}} \pubyear{2016}

\begin{document}

\label{firstpage}
\maketitle

\begin{abstract}
Molecular clouds are to a great extent influenced by turbulent motions in the gas. Numerical and 
observational studies indicate that the star formation rate and efficiency crucially depend on the mixture of solenoidal and compressive 
modes in the turbulent acceleration field, which can be quantified by the turbulent driving parameter $b$. For purely solenoidal 
(divergence--free) driving previous studies showed that $b=1/3$ and for entirely compressive (curl--free) driving $b=1$.
In this study, we determine the evolution of the turbulent driving parameter $b$ in magnetohydrodynamical simulations of molecular cloud formation and evolution. The clouds form due to the convergence of two flows of warm neutral gas.
We explore different scenarios by varying the magnitude of the initial turbulent perturbations in the flows. We show that the driving mode of the turbulence within 
the cloud strongly fluctuates with time and exhibits no clear correlation with typical cloud properties, such as the cloud mass and the (Alfv\'{e}n) Mach number. We specifically find that $b$ strongly 
varies from $b\sim0.3$ to $b\sim 0.8$ on timescales $t\lesssim5\,$Myr, where the timescale and range of variation can change from cloud to cloud. This rapid change of $b$ from solenoidal to compressive driving is primarily associated with 
global contraction of the cloud and subsequent onset of star formation. We conclude that the effective turbulence driving parameter should be treated as a free parameter that can vary from solenoidal to compressive in both time and space.
\end{abstract}
\section{Introduction}
Turbulence plays a key role in astrophysics and is of utmost importance for understanding the formation of stars in galaxies \citep[][]{Elmegreen04,Padoan14}. Its role is two--fold.
Turbulent fluctuations 
may disperse the gas due to the random nature of the (turbulent) velocity field. In this sense, turbulence provides a form of support against gravity. On the other hand, 
 observations and numerical simulations have shown that interstellar turbulence is supersonic \citep[e.g.][]{MacLow04,McKee07,Federrath08,Schmidt13}. Hence, turbulent energy can be dissipated in shocks and the resulting compression might form stable 
overdensities, which eventually undergo gravitational collapse. Furthermore, the supersonic interstellar turbulence must be driven by some external stirring mechanism, as otherwise, it would decay within a crossing--time \citep[][]{MacLow98a,Stone98,Ostriker99,MacLow04}. \\
Possible driving agents of turbulence are gravity, galactic accretion and shear, the magneto--rotational instability, cloud--cloud collisions and colliding flows, as well as stellar feedback \citep[][]{MacLow04,Scalo04,Federrath10b,Federrath12,Federrath16d,Federrath17}. The major difference between those driving mechanisms is the way \ita{how} they drive the 
turbulence -- either solenoidal (divergence--free) or compressive (curl--free). The driving mechanism thus determines the nature of the velocity field in the gas and 
hence either promotes or hinders the gas from forming large overdensities. However, due to the non--linear nature inherent to turbulence both compressive and solenoidal motions will be excited with time. Furthermore, the compressive modes decay faster and hence, on a certain spatial scale, there will be dominant solenoidal modes 
even if the driving is almost entirely compressive \citep{Federrath09a,Federrath10b,Federrath13}.\\
The density probability distribution function (PDF) of isothermal, turbulent gas was found to be approximately log--normal \citep[][]{Passot98,Vazquez00b}. The width of the PDF is determined by 
the turbulent Mach number, the plasma--$\beta$ and the driving parameter, $b$. Turbulence stirring can be quantified by the latter parameter, which smoothly varies between 1/3 and 1, where these two extreme cases refer to purely solenoidal and entirely compressive driving \citep[][]{Federrath10,Herron17}. 
As was shown by \citet[][]{Federrath08}, the flow dynamics and the density structure strongly depend on whether turbulence is driven in a  solenoidal or compressive way. Purely compressive driving results in a broader density PDF and thus in a larger fraction of gas at high densities. In contrast, solenoidal driving with $b=1/3$ shows smaller 
widths of the resulting PDF, because primarily divergence--free/rotational modes are excited which prevent the build--up of large overdensities.
As was shown by \citet[][]{Federrath12}, compressive driving thus results in a 
star formation rate approximately an order of magnitude larger than for solenoidal driving. In addition, the authors showed that the resulting star formation efficiency in turbulent molecular clouds can be directly linked to the slope of the PDF at high densities \citep[][]{Federrath13}.\\ 
The star formation rate and efficiency in molecular clouds strongly depend on properties, which are largely governed by turbulence, i.e. the turbulent Mach number 
$\mathcal{M}$, the virial parameter $\alpha_\mathrm{vir}$, the turbulent driving parameter $b$ and the density variance $\sigma_{\varrho/\varrho_0}^2$. For 
example, in certain models the star formation rate per free--fall time, that is, the amount of stars formed within a cloud's free--fall time, takes the general (multi--free--fall) form \citep{Federrath12}
\beq\label{eqSFR}
\mathrm{SFR_{ff}} = \frac{\epsilon}{2\phi_\mathrm{t}}\mathrm{exp}\left(\frac{3}{8}\sigma_s^2\right)\left[1+\mathrm{erf}\left(\frac{\sigma_s^2-s_\mathrm{crit}}{\sqrt{2\sigma_s^2}}\right)\right].
\eeq
Here, $s=\mathrm{ln}\left(\varrho/\varrho_0\right)$ is the logarithmic density contrast, $\epsilon$ parameterises the fraction of material re--injected via stellar feedback \citep[see e.g.][]{Federrath14} and $1/\phi_\mathrm{t}$ is the uncertainty in the multi--free--fall ansatz, but generally 
$1/\phi_\mathrm{t}\sim1$. The density variance is a function of the Mach number, the plasma--$\beta$ and the driving parameter, whereas the critical density, $s_\mathrm{crit}$, depends in addition on the virial ratio $\alpha_\mathrm{vir}$. The exact form of $s_\mathrm{crit}$ depends on the model and the assumptions used \citep{Krumholz05c,Padoan11,Hennebelle11,Federrath12}. Due to the dependence of $s_\mathrm{crit}$ and $\sigma_s$ on the above mentioned parameters, slight variations in these will have a large impact on the 
resulting SFR of the clouds \citep[see e.g., the importance of the turbulence driving parameter for the Central Molecular Zone cloud G0.253+0.016][]{Federrath16d}.\\
Observationally, the type of turbulence driving in molecular clouds is hard to disentangle because of the need for high spatial and spectral resolution in order to measure meaningful density PDFs and 
line data. A major issue in comparing to observations is also that one can only measure the column density PDF, while the theoretical models are based on the volume density PDF. However, recent work based on statistical arguments have derived methods to overcome this problem \citep[][]{Brunt10b,Brunt10c,Brunt14,Kainulainen14}.Recent studies on the 
driving of turbulence in nearby clouds revealed that it is rather compressive with 
$b\gtrsim0.4$ \citep[][]{Padoan97,Brunt10,Ginsburg13}. In contrast, \citet[][]{Kainulainen13b} argue that numerical 
simulations with a magnetic field of the order of $B\sim3\,\mu\mathrm{G}$ and rather solenoidal driving, $b\lesssim0.4$, fit 
best observations of solar--neighbourhood clouds presented in \citet[][]{Kainulainen13a}. More recently, \citet[][]{Federrath16d} 
derived $b=0.22\pm0.12$ for the central molecular zone cloud G0.253+0.016, indicating pre--dominantly 
solenoidal driving in the Galactic center environment.\\
In this study we focus on the evolution of the turbulent driving parameter in simulations of molecular cloud formation and evolution. After the 
flows have deceased, the 
turbulence within the clouds is solely driven by gravity and we will show that $b$ is subject to large temporal variations.\\
In section \ref{secNumerics} we briefly introduce the numerical 
model and explain the method of deriving $b$. In section \ref{secResults} we discuss our findings. This study is closed by a discussion of limitations in section \ref{secLimits} and a summary in section \ref{secSummary}.
\section{Methodology}\label{secNumerics}
\subsection{Numerical Simulations}
The numerical details of the simulations presented in this study have already been discussed in detail in \citet[][henceforth K16]{Koertgen16}.\\
 The simulations are performed using the FLASH code\footnote{http://flash.uchicago.edu} (v2.5). Two flows of warm neutral medium (WNM) gas collide head--on in the center of the simulation box. Each flow contains approximately 50,000\,M$_\odot$ and is 
 supersonic with respect to the WNM with \ita{isothermal} Mach number $\mathcal{M}_\mathrm{flow}=2$. 
 Additionally, the flows are  
turbulent with RMS Mach numbers ranging from $\mathcal{M}_\mathrm{RMS,flow}=0.8$ to $\mathcal{M}_\mathrm{RMS,flow}=1.2$. The flows are initially magnetically critical with 
$\mu/\mu_\mathrm{crit}\sim 1\,\left(B_0=3\,\mu\mathrm{G}\right)$. 
We use 11 levels of refinement, giving a maxi\-mum spatial resolution of $\Delta x=0.03\,\mathrm{pc}$. Sink particles are included in order to replace collapsing, star--forming regions, which otherwise would result in 
under--resolving the local Jeans length of the densest parts of the cloud \citep[e.g.][]{Truelove97, Federrath10}. The 
simulations are summarised in Table \ref{tabIC}.
\subsection{Measuring the turbulent driving parameter $b$}
\begin{figure}
\centering
\includegraphics[width=0.4\textwidth,angle=-90]{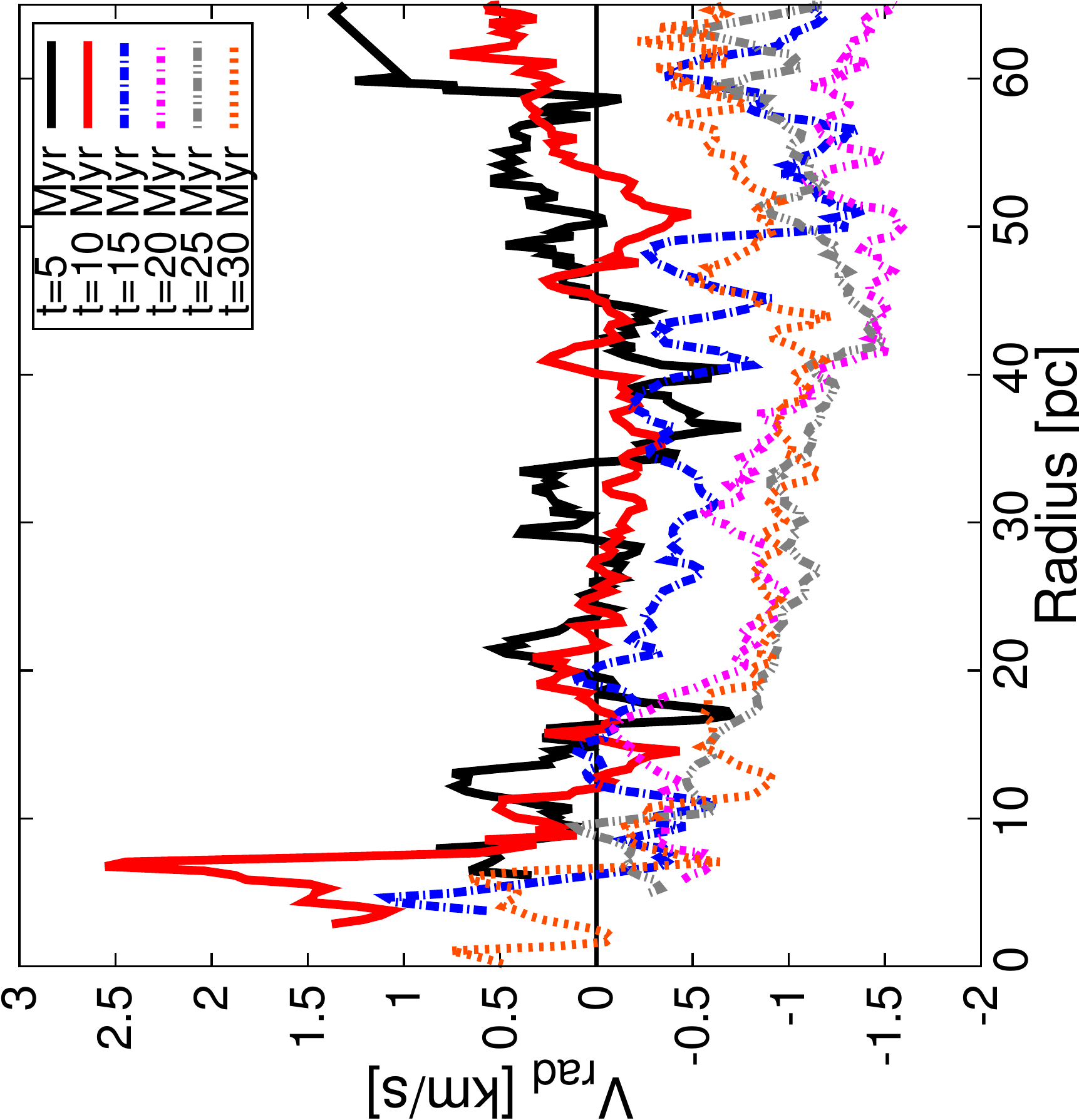}
\caption{Radial profile of the contraction velocity at different times for run MHD--M1.0. Positive velocities indicate 
expansion at this radius, negative velocities contraction towards the gravitational potential well. As can be seen, 
at early times there is a mixture of expansion and contraction. At later times, the majority of the cloud is in a state of collapse and a gradient in the velocity profile is established, although this gradient is rather shallow.}
\label{figRad}
\end{figure}

To derive the driving parameter $b$, we use the density PDF. The dispersion, $\sigma_{\varrho/\varrho_0}$, of the PDF depends on the driving parameter, the turbulent Mach number, $\mathcal{M}=\sigma_\mathrm{v}/c_\mathrm{s}$, and the 
ratio of thermal to magnetic pressure $\beta=2c_\mathrm{s}^2/v_\mathrm{a}^2$ via \citep[][]{Federrath08,Padoan11a,Molina12,Nolan15,Federrath15}
\beq
\sigma_{\varrho/\varrho_0}\,=\,b\mathcal{M}\frac{1}{\sqrt{1+\beta^{-1}}}.
\label{eqSigma}
\eeq
We note that the density PDF does not necessarily have to be log--normal for  Eq. $\left(\ref{eqSigma}\right)$ to be applicable and one is only faced with solving for $b$. The appearing values for $\beta$ and $\mathcal{M}$ are   
volume--weighted averages. Eq. $\left(\ref{eqSigma}\right)$ is evaluated only for the gas with $n=10^2-10^4~\mathrm{cm}^{-3}$. We take the lower limit as the threshold density to assume the gas to be molecular and part of the cloud \citepalias[see also][]{Koertgen16}. The upper limit is chosen such that regions in 
the immediate vicinity of sink particles are excluded as the velocity field in these is dominated by ordered accretion flow onto the particles rather than turbulence. The fluctuations in this high--density regime do not significantly alter the average turbulent Mach number. However, the density in these regions still increases thereby biasing the variance of the density--PDF, which leads to an artificially increasing ratio of $\sigma_{\varrho/\varrho_0}/\mathcal{M}$.\\
At late times, the clouds undergo global contraction, preferentially in the radial direction \citep[][]{Vazquez07}. To 
extract turbulent motions, we calculate radial velocity profiles as well as velocity profiles \ita{along} the flow direction and subtract the average contraction velocities 
at each radius and 'height' (see Fig. \ref{figRad}). The Mach number, which enters Eq. $\left(\ref{eqSigma}\right)$, thus contains only the turbulent fluctuations. We further calculate and remove radial density 
gradients, which otherwise bias the derivation of the density dispersion due to a global collapse--profile \citep[see][]{Federrath11a,Padoan16,Pan16}.\\
The error $\Delta b$ of the driving parameter is calculated according to error propagation, where we assume variation in the sound and Alfv\'{e}n speed, 
respectively.
\begin{table}
\caption{Overview of the simulations presented in this study.} 
\begin{tabular}{cccc}
\hline
\hline
\fat{Run}	&\fat{Min. $\Delta\mathrm{x}$}	&\fat{$\mathcal{M}_\mathrm{RMS,flow}$}		&\fat{Sim. End}\\
			&$\left(\mathrm{pc}\right)$	&	&$\left(\mathrm{Myr}\right)$ \\
MHD--M0.8	&0.03	&0.8	&28 \\
MHD--M1.0	&0.03	&1.0	&30 \\
MHD--M1.2	&0.03	&1.2	&33 \\
\hline
\hline
\end{tabular}
\label{tabIC}
\end{table}
\subsection{Resolution study}
\begin{figure}
\centering
\includegraphics[width=0.4\textwidth,angle=-90]{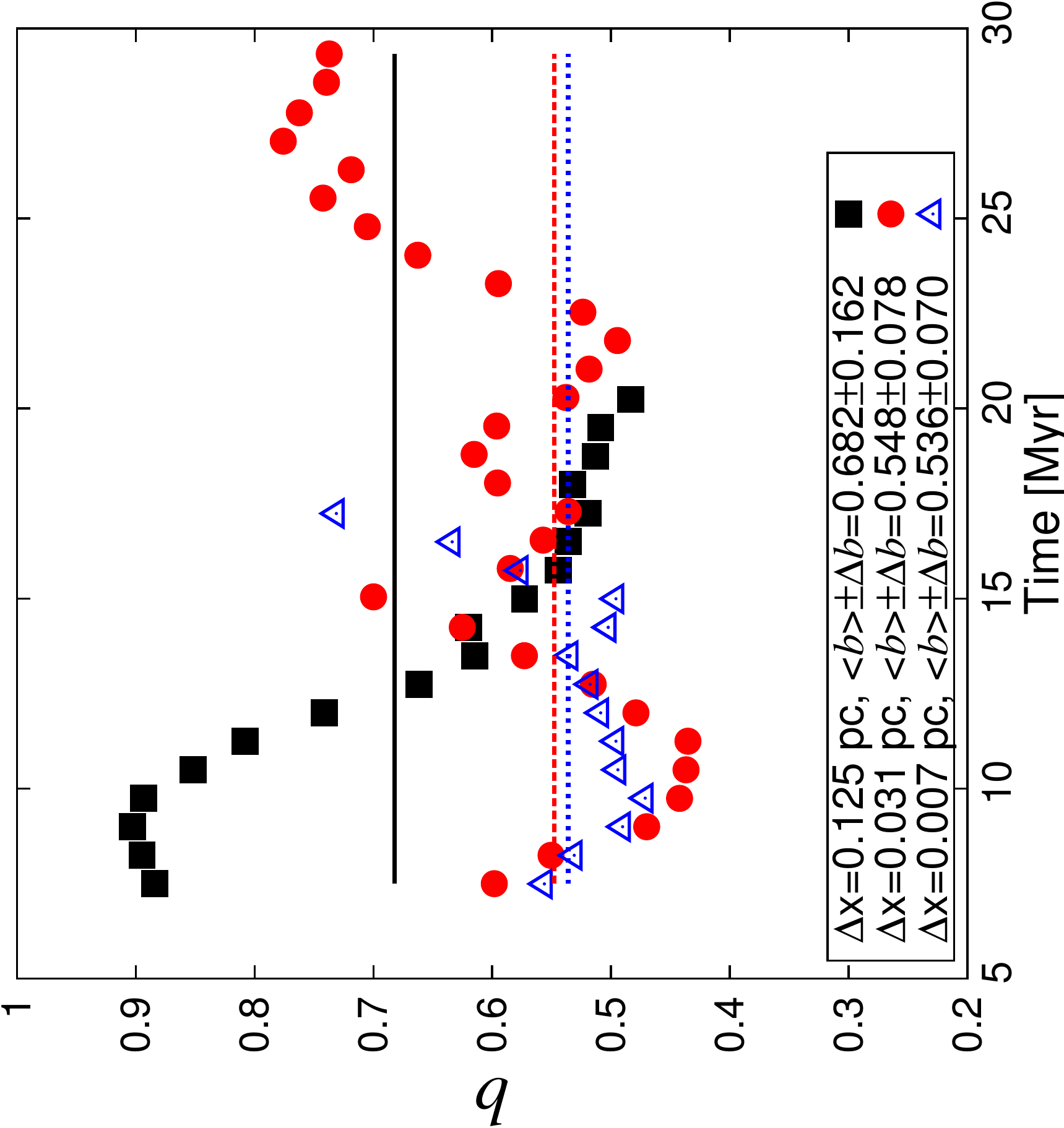}
\caption{Resolution study showing the convergence of the simulations with increasing numerical resolution. Depicted is the average \bp as a function of time. The horizontal lines denote the time--averaged \bp with the respective value as well as the standard deviation given in 
the plot legend. For reasons of comparability the averaging is performed for times $t\in\left[5,20\right]\,\mathrm{Myr}$. Although all simulations show periods of strong variation, the average is converged. To achieve convergence, we recommend a grid resolution of  $\Delta\mathrm{x}\lesssim0.1\,\mathrm{pc}$.}
\label{figResstudy}
\end{figure}

The decay rate of the turbulent energy and the general properties of turbulence in numerical simulations depend on the resolution adopted with lower resolution showing e.g. a larger (artificial) decay \citep[][]{MacLow98a,Federrath09a}. In order to ensure convergence of our results, we solve Eq. $\left(\ref{eqSigma}\right)$ for $b$ for different numerical resolutions. Fig. \ref{figResstudy} shows the average driving parameter as a function of time for the cloud formed in run MHD--M0.8. The different colours depict 
different numerical 
resolutions, namely $\Delta\mathrm{x}=0.1\,\mathrm{pc}$ (black squares), $\Delta\mathrm{x}=0.03\,\mathrm{pc}$ (red dots) and $\Delta\mathrm{x}=0.007\,\mathrm{pc}$ (blue open triangles), respectively. The lines denote the time--averaged value of $b$, which is given 
in the figure legend including its standard deviation. The temporal evolution shows strong variation of \bp with time.
However, the average value is converged at resolutions $\Delta\mathrm{x}\lesssim0.1\,\mathrm{pc}$. 
\section{Results}\label{secResults}
\subsection{The density probability distribution function of the clouds}
\begin{figure}
\centering
\includegraphics[width=0.4\textwidth,angle=-90]{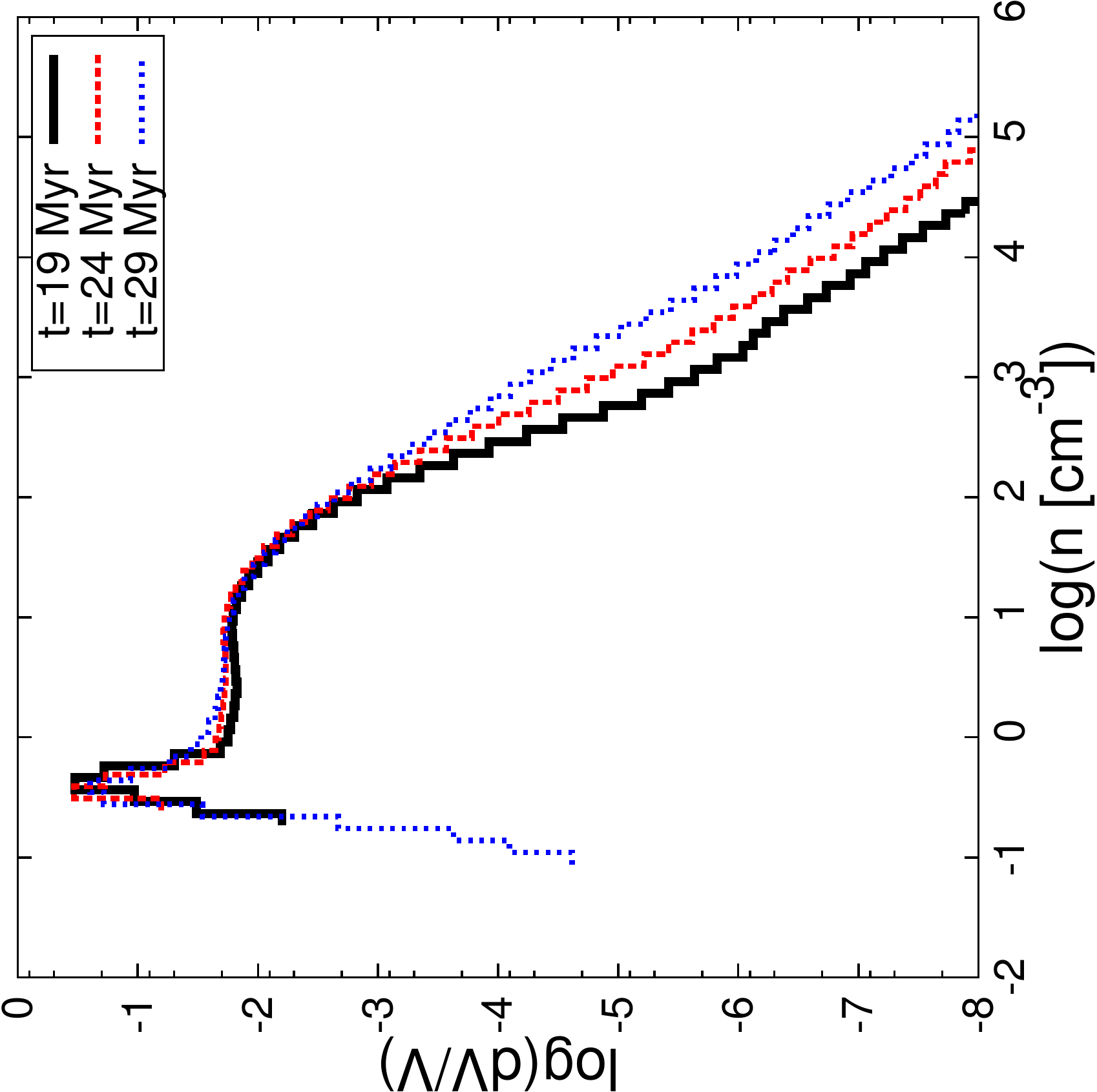}
\caption{Number density PDF at three stages in run MHD--M0.8 covering a time span of $\Delta t\,=\,10\,\mathrm{Myr}$. The PDF is clearly divided into three parts. 1) A narrow log--normal part at $\mathrm{log}\left(n\right)\in\left[-1,0\right]$, 2) a broader 
log--normal distribution at around $n\sim10-10^3\,\mathrm{cm}^{-3}$ and 3) a power--law tail at high densities. The power--law tail is an indication of 
collapsing gas. Note the flattening of the slope with time.}
\label{figPDF}
\end{figure}

The complex interplay of gravity, MHD turbulence and thermodynamics shapes the resulting density PDF. Fig. \ref{figPDF} shows the density distribution in run MHD--M0.8 at different times. 
The total PDF consists of two peaks, which depict the contributions from the WNM and the cold neutral medium (CNM). Both contributions are close to log--normal. At high gas 
densities, self--gravity produces a power--law tail as gas is gradually collapsing, which is also seen in column density PDFs of nearby star forming clouds \citep[][]{Kainulainen09,Schneider13,Schneider14,Schneider16}. With time the PDF power--law tail 
begins to flatten due to the increased amount of gas in the gravitationally unstable regime \citep[][]{Federrath13}. The 
formed molecular clouds are defined to consist of gas with $n\geq100\,\mathrm{cm}^{-3}$. Hence, the clouds themselves 
are described by a dominant power--law tail and some portion of the log--normal part of the CNM gas distribution.

\subsection{Evolution of the turbulent driving parameter}
\subsubsection{Dependence on cloud properties}
\begin{figure*}
\begin{tabular}{ccc}
\includegraphics[width=0.29\textwidth,angle=-90]{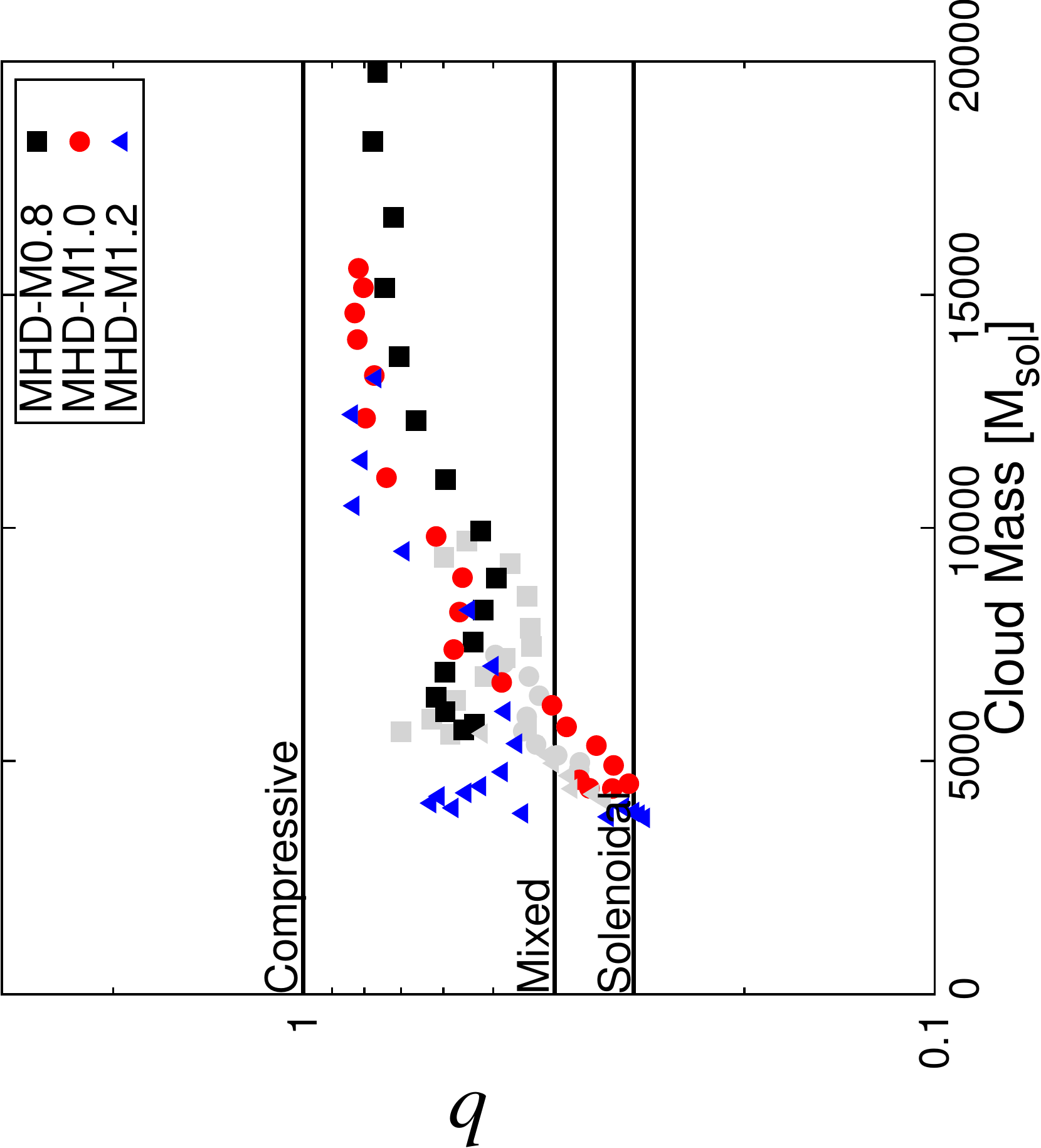}&\includegraphics[width=0.287\textwidth,angle=-90]{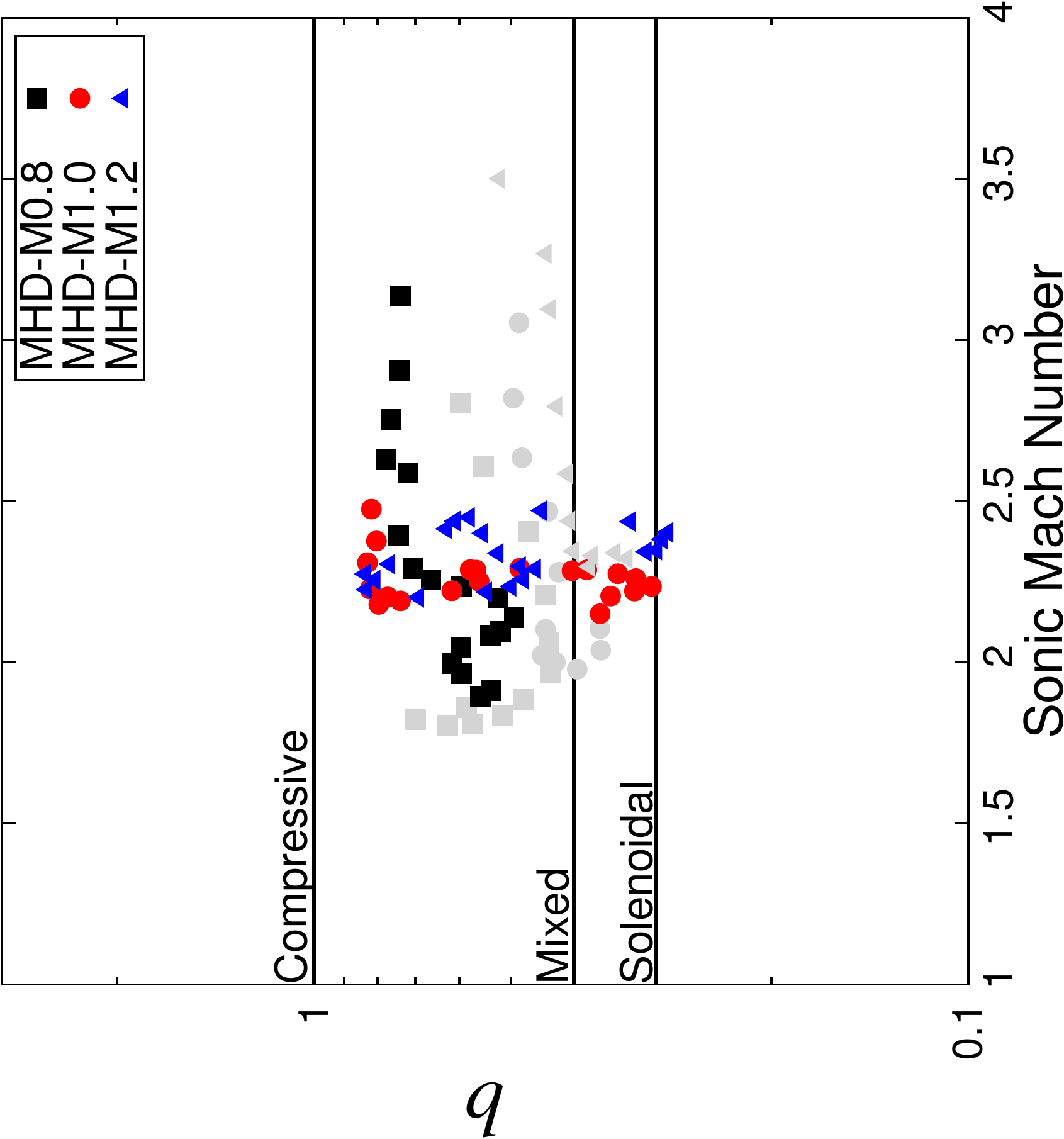}
\includegraphics[width=0.29\textwidth,angle=-90]{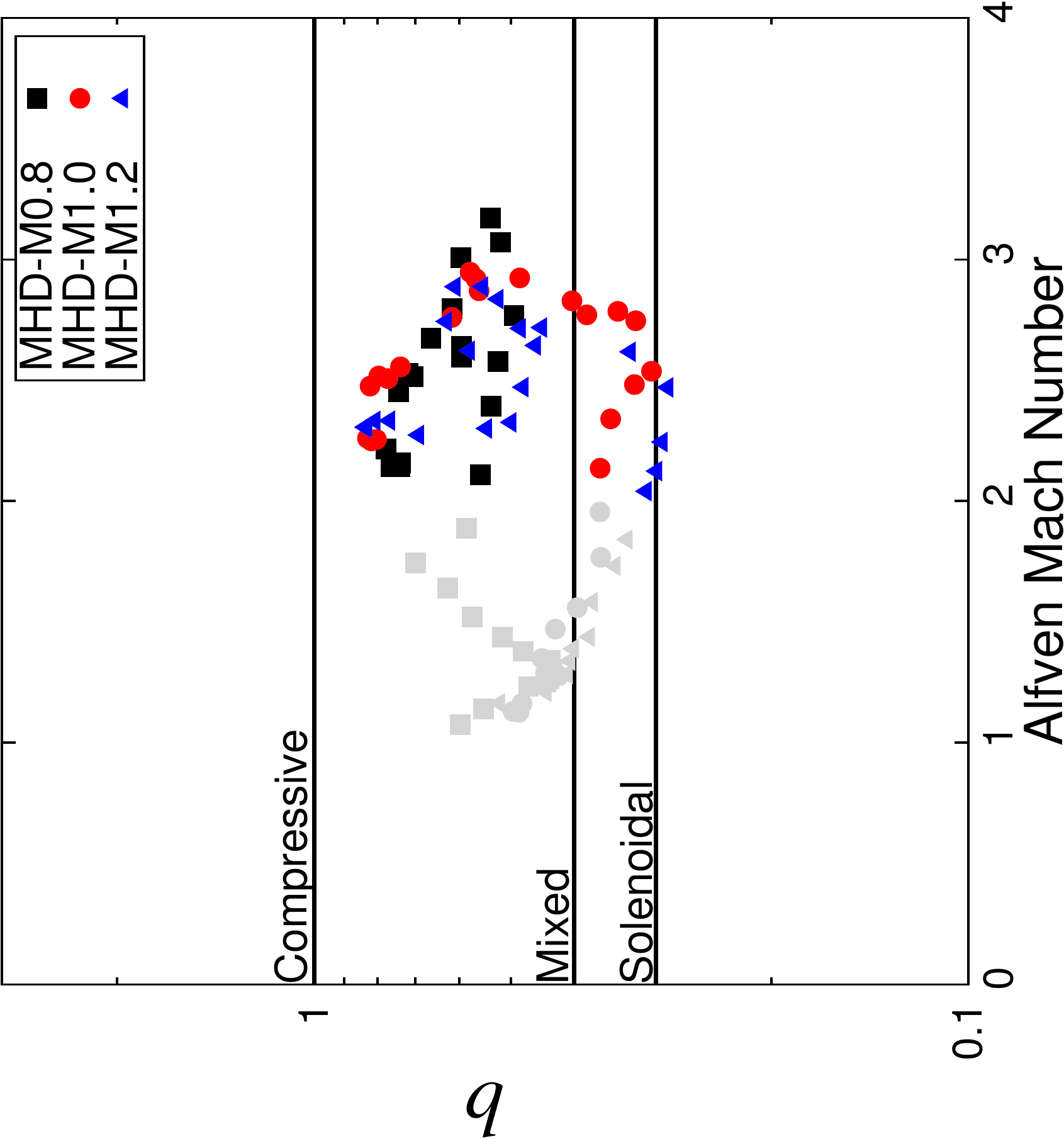} 
\end{tabular}
\caption{The driving parameter $b$ as a function of various cloud properties. The grey shaded data points have $\mathcal{M}_\mathrm{A}<2$. \ita{Left:} \bp as a function of cloud mass. At small cloud masses there appears to be no correlation between $b$ and the mass. By contrast, with increasing cloud mass, the turbulence
becomes more compressive (i.e. $b\rightarrow1$). \ita{Middle:} \bp as a function of the turbulent Mach number of the cloud. In the range of Mach numbers probed $b$ is independent 
of the Mach number. \ita{Right:} \bp as a function of the cloud's turbulent Alfv\'{e}n Mach number. $b$ shows a great spread as function of $\mathcal{M}_\mathrm{A}$, but again, obvious correlations do not emerge. }
\label{figbprop}
\end{figure*}

Fig. \ref{figbprop} shows \bp as a function of various cloud properties for the different simulations. Grey symbols indicate $\mathcal{M}_\mathrm{A}<2$, where
$\mathcal{M}_\mathrm{A}=\sigma_v/v_\mathrm{A}$ is the turbulent Alfv\'{e}n Mach number. As stated in \citet[][]{Molina12}, $\mathcal{M}_\mathrm{A}<2$ indicates a significant contribution from the magnetic field, which imposes anisotropy to the fluid flow. The derivation of Eq. \ref{eqSigma}, however, 
assumes the turbulence to be fully isotropic and hence the applicability of Eq. \ref{eqSigma} is questionable in this regime of Alfv\'{e}n Mach numbers and is 
thus omitted (this data is shown in light grey).\\
We find that \bp takes a variety of values as function of cloud mass (left), ranging from purely solenoidal to purely compressive driving. The scatter in \bp is larger at lower cloud masses, 
which depict the initial stages of the clouds formed in between the flows where the latter are affected by the initial turbulent fluctuations. At higher masses there appears to be a trend of increasing $b$ with increasing cloud mass. At these stages, which correspond to times $t>20\,\mathrm{Myr}$, the initial turbulent fluctuations have decayed and the externally compressing flows have vanished. The dynamical evolution of the clouds is then entirely controlled by gravity, turbulence 
and magnetic fields. Once, the clouds are magnetically super--critical and sub--virial, gravitational contraction leads to a highly compressive velocity field which 
promotes the formation of gravitationally unstable overdensities. \\
Turbulence is not externally driven in the simulations presented here. The resulting range of Mach numbers is thus relatively narrow and we do not find any 
trend of $b$ as function of Mach number. At small Mach numbers there appears a large scatter in $b$, again referring to early times where the converging 
flows stir up the dense gas. At slightly higher Mach numbers, $b$ stays almost constant. From Eq. \ref{eqSigma} there should occur a decrease in $b$ with 
increasing Mach number, which is not observed here. As $\sigma_{\varrho/\varrho_0}$ is not kept constant, it is evident that the density dispersion significantly increases as the Mach number increases (which is expected), thereby keeping the ratio $\sigma_{\varrho/\varrho_0}/\mathcal{M}$ approximately constant.\\
In the right plot of Fig. \ref{figbprop} we show $b$ as a function of the Alfv\'{e}n Mach number. Again, $b$ shows a large scatter and no clear correlation with the 
cloud's Alfv\'{e}n Mach number, which might be attributed to the small range of Mach numbers probed. However, the scatter is larger for clouds with an initially 
higher turbulent Mach number within the flows. \\
To summarise, $b$ appears to be nearly independent of the cloud properties discussed in this section. Only for the cloud mass a trend can be recognised, but 
we caution that the probed mass range is rather narrow. Besides these findings it is seen that $b$ is on average larger, i.e. pointing towards more compressive 
turbulence, for lower initial Mach numbers within the converging WNM flows. Whether this trend holds for different Mach numbers and larger cloud masses 
needs to be investigated in the future. 
\subsubsection{Time evolution}
\begin{figure}
\includegraphics[width=0.45\textwidth,angle=-90]{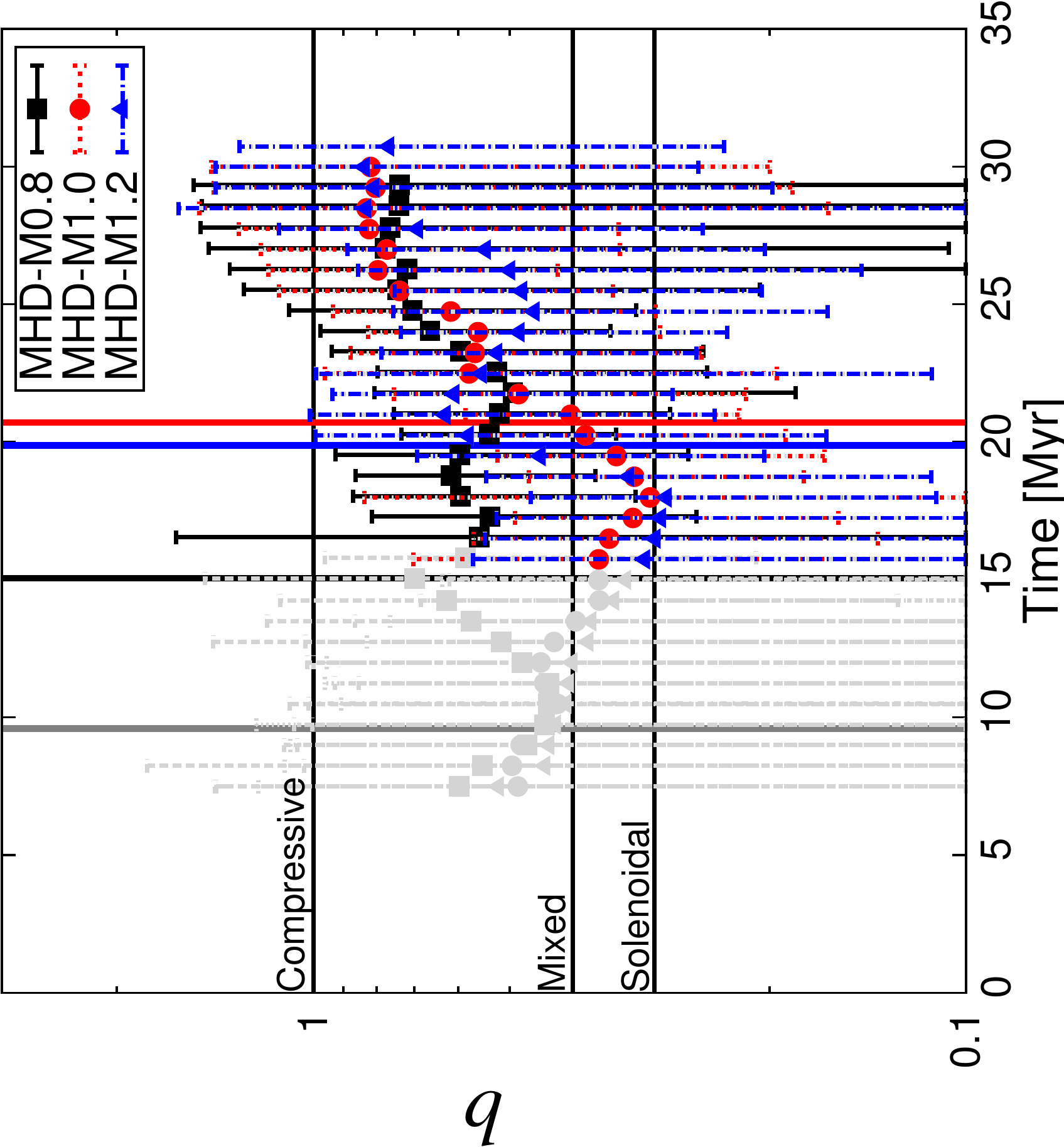}
\caption{Time evolution of the turbulent driving parameter for different initial conditions. Grey shaded data have $\mathcal{M}_\mathrm{A}<2$. The dark grey 
vertical line at $t=9.6\,\mathrm{Myr}$ gives the dynamical time of the converging WNM flows. The black, blue, and red vertical lines, respectively, denote the time 
at which the first sink particle has formed in the individual clouds. For trans-- to supersonic initial fluctuations in the flows the driving parameter is $b\lesssim0.4$ for $t\lesssim20\,\mathrm{Myr}$, whereas it is rather compressive for subsonic fluctuations. When sink particles are formed, all clouds show more 
compressive turbulence driving. However, $b$ still varies by some amount before it starts to approach a constant value of $b\sim0.8$ from $t\sim27\,\mathrm{Myr}$ on. This indicates that, in case of the cloud in run MHD--M1.2, gravitational contraction is accompanied by fragmentation of the collapsing gas. 
Note that the uncertainty in $b$ encloses the whole range of allowed values. Solenoidal, mixed, as well as compressive driving are indicated by the 
horizontal black lines.\newline (For a colour version see online manuscript.)}
\label{figTime}
\end{figure}

In the previous section we showed that there only exists a weak correlation of \bp with different properties of the molecular cloud and that it strongly varies. In this section we focus on the 
temporal evolution of the driving parameter. The results are shown in Fig. \ref{figTime}. Grey symbols again indicate data where $\mathcal{M}_\mathrm{A}<2$. The 
grey vertical line indicates the dynamical time of the WNM flows. The black, blue, and red lines, respectively, depict the times when the first sink particle has formed 
in the individual clouds.\\
At early stages, $t<10\,\mathrm{Myr}$, \bp increases primarily due to the compression by the WNM streams. Since the magnitude of the fluctuations in the streams is much weaker than the bulk velocity, the energy in the compressive modes is larger 
than in the solenoidal modes. However, we note that this is biased by the choice of our initial conditions. Once the clouds start expanding due to the lack of external pressure by the WNM flows, \bp decreases again as now the re--expanding gas interacts non--linearly with the gas in the diffuse halo surrounding the cloud, which is about to be accreted. By increasing the 
magnitude of the turbulent fluctuations within the flow, the average \bp is 
decreased due to less efficient compression at the cloud outskirts. However, also in the clouds with higher turbulent Mach numbers, \bp is still more likely to be compressive, i.e. $b\sim0.4-0.5$\footnote{We caution that these data are biased by $\mathcal{M}_\mathrm{A}<2$ where our analysis is not applicable as stated by \citet[][]{Molina12}.}.  \\
From $t\sim15\,$Myr on the turbulence becomes super--alfv\'{e}nic with $\mathcal{M}_\mathrm{A}>2$. The trend of decreasing $b$ with time is continued for 
runs MHD--M1.0 and MHD--M1.2, where the turbulence has now become almost solenoidal. In contrast, the cloud in run MHD--M0.8 shows $b\sim0.6$, 
indicating rather compressive driving. Note that at this time a sink particle has already formed in the latter run, which is indicative of highly gravitationally unstable 
regions in this cloud. At $t\sim20\,$Myr, $b$ increases also for the higher turbulent clouds due to the onset of gravitational collapse, which results in sink particle formation. However, the clouds still undergo 
fragmentation after the formation of sink particles, which explains the fluctuations in $b$ with time. At $t\sim30\,\mathrm{Myr}$, $b$ is almost identical in all clouds, which are now forming sink particles. The variation in $b$, 
as given by the error bars, indicates that, though the clouds show on average a compressive turbulent velocity 
field, there are regions inside the clouds, which exhibit almost entirely solenoidal driving, typical for a globally 
contracting and locally fragmenting cloud.
\section{Limitations}
\label{secLimits}
Our analysis is limited due to assumptions made or due to conditions which arose during the simulations. \\
First, the calculated turbulent Mach numbers range only up to around $\mathcal{M}_\mathrm{RMS}\sim4$. Although the simulated Mach numbers 
are rather low, they do still fit with observations of e.g. Orion B and other molecular clouds \citep[][]{Schneider13}. We are thus only able to trace some part of the evolution of $b$ as function of the Mach number. Simulations that probe a wider range of Mach numbers and cloud masses need to be investigated in the future. \\
Second, the $\sigma_{\varrho/\varrho_0}-b$ relation is sensitive to the scaling of the magnetic field with density, as was pointed out by \citet[][]{Molina12}. In our study, we ignored possible variations and focussed on a scaling similar to what has been 
derived from numerical simulations of colliding flows, namely $B\propto\varrho^{1/2}$ \citep[e.g.][]{Koertgen15}. A slope taken from observations \citep[e.g. $B\propto\varrho^{2/3}$,][]{Crutcher10,Crutcher12} should not significantly impact our analysis. This is also 
because the magnetic field -- density relation has a large spread in (simulated) molecular clouds \citep[][]{Banerjee09a,Molina12,Koertgen15}.\\
Third, we do not fit a Gaussian to the obtained density PDF and might under--estimate the width of the PDF. \\
Here we restricted our analysis to simulations without feedback and focussed on the global cloud properties and their relation to the driving of turbulence. Although our resulting $b$ agrees on average well with observational constraints for non--star forming clouds by \citet[][]{Ginsburg13}, stellar feedback through jets/outflows, winds and/or 
supernovae should influence the evolution of $b$. However, we are still able to probe the evolution of $b$ in the absence of 
feedback, which enables us to study the effect of pure dynamics. To what extent the feedback will influence $b$, however, is currently not clear and will be investigated in a subsequent study based on simulations by \citet{Koertgen16}.
\section{Summary}
\label{secSummary}
We have presented a set of high--resolution simulations of the formation and evolution of molecular clouds formed by colliding streams of warm neutral gas. The focus of this study was the analysis of 
the turbulent driving parameter $b$. For this purpose we analysed the relation between the volume--density dispersion and the type of driving in star--forming molecular clouds without any form of stellar feedback. We find that $b$ can vary between solenoidal and compressive driving from cloud to cloud and can rapidly change on timescales of only a few Myr even within a single cloud. The most 
prominent change is primarily associated with the onset of sink particle formation and global contraction of the cloud. Prior to sink formation we find $b\sim0.3-0.5$, whereas $b\sim0.8$ at late times.
Although it appears that $\left<b\right>\sim0.5$ over the course of the cloud evolution, which would indicate rather compressive forcing, we lastly conclude that there is no fixed mode of turbulent driving.

\section*{Acknowledgements}
B.K.~and R.B.~thank Rachid Ouyed for useful discussions on MHD turbulence. B.K. further thanks Jouni Kainulainen for inspiring discussions. B.K.~and R.B.~acknowledge funding from the German Science Foundation (DFG) within the Priority Programm "The Physics of the ISM" (SPP 1573) via the grant BA 3706/3-2. R.B.~further 
acknowledges funding for this project from the DFG via the grants BA 3706/4-1, BA 3706/14-1 and BA 3706/15-1.
C.F.~acknowledges funding provided by the Australian Research Council's Discovery Projects (grants~DP150104329 
and~DP170100603). C.F.~thanks for high performance computing resources provided by the Leibniz Rechenzentrum 
and the Gauss Centre for Supercomputing (grants~pr32lo, pr48pi and GCS Large-scale project~10391), the Partnership 
for Advanced Computing in Europe (PRACE grant pr89mu), the Australian National Computational Infrastructure 
(grant~ek9), and the Pawsey Supercomputing Centre with funding from the Australian Government and the Government 
of Western Australia, in the framework of the National Computational Merit Allocation Scheme and the ANU Allocation 
Scheme. The simulations were run on HLRN--III under project grand hhp00022. The software used in this work was in 
part developed by the 
DOE--supported ASC/Alliance Center for Astrophysical Thermonuclear Flashes at the University of Chicago.
\bibliography{astro}
\bibliographystyle{mn2e}
\end{document}